\documentstyle[prc,preprint,aps,epsf]{revtex}

\begin{document}
\draft
\title{NN SCATTERING: CHIRAL PREDICTIONS FOR ASYMPTOTIC OBSERVABLES}
\author{J-L. Ballot$^1$\footnotemark[1]}
\address{Division de Physique Th\'eorique, Institut de Physique
Nucl\'eaire$^2$\footnotemark[2]\\ F-91406, Orsay CEDEX, France}
\author{ M.R. Robilotta$^3$\footnotemark[3]}
\address{FINPE, Instituto de F\'{\i}sica, Universidade de S\~ao Paulo \\
C.P. 66318, 05389-970, S\~ao Paulo, SP, Brasil}
\author{C.A. da Rocha$^4$\footnotemark[4]}
\address{Department of Physics, University of Washington\\
Box 351560, Seattle, Washington 98195-1560, U.S.A.}
\date{March, 1997}
\maketitle

\footnotetext[1]{$^{*}$Electronic address: ballot@ipno.in2p3.fr}
\footnotetext[2]{$^{\dag}$Unit\'e de Recherche des Universit\'es Paris 6 
et Paris 11 associ\'e au CNRS.}
\footnotetext[3]{$^{\ddagger}$Electronic address: robilotta@if.usp.br}
\footnotetext[4]{$^{\S}$Present address:  Instituto de F\'{\i}sica Te\'orica (IFT), 
Universidade Estadual Paulista (UNESP) - R. Pamplona, 145 - 01405-900 - 
S\~ao Paulo, SP, Brazil. Electronic address: carocha@ift.unesp.br}

\begin{abstract}
We assume that the nuclear potential for distances larger than 
2.5 fm is given just by the exchanges of one and two pions and, for the
latter, we adopt a model based on chiral symmetry and subthreshold
pion-nucleon amplitudes, which contains no free parameters.
The predictions produced by this model 
for nucleon-nucleon observables
are calculated and shown to agree well
with both experiment and
those due to phenomenological 
potentials.
\end{abstract}


\newpage
\section{INTRODUCTION}

In the last two decades several semiphenomenological potentials 
were proposed \cite{S73,d75,P80,A84,B87} which can describe rather 
well NN scattering data below the pion production threshold.  All 
these models have a common feature, namely that the interaction at 
large distances is ascribed to the one-pion exchange 
potential (OPEP). On the other hand, they differ significantly 
at intermediate and short distances,
a fact that becomes evident when one inspects the profile functions
produced for the various components of the force. In each case, the 
reproduction of experimental data is achieved by means of a different
balance between effects due to long and short distances. The fact that
these potentials are successful means that somehow they are able to 
incorporate the relevant average dynamics.

At present, none of the existing semiphenomenological potentials
include explicitly the dynamics associated with chiral symmetry, which
constitutes the main conceptual framework for the study of strong
interactions at energies which are small compared to the QCD scale. In
this regime, non-perturbative effects are dominant and one is not able
to do calculations using QCD directly.
The usual strategy for overcoming this problem
consists in working with an effective theory, constructed in such
a way as to include as much as possible the main features of the basic theory.
The masses of the quarks $u$ and $d$ are very small and hence their
interactions with gluons are approximately invariant under the group 
$ SU(2) \times SU(2) $. Therefore, one requires the effective theory
at the hadron level posses the same basic symmetry, broken by just
the pion mass. 

In the last five years several authors have tackled the 
problem of NN interactions in the light of chiral symmetry
and, so far, only processes associated with the two-pion
exchange potential (TPEP) were systematically studied
\cite{OK92,CPS92,FC94,RR94,B94,ORK94,RR95,R95,ORK96,RR97}.
Chiral symmetry is very relevant to this component of the
force because it controls the behaviour of the intermediate
$\pi$N amplitude, that is the main building block of the
interaction.
The first works of this series were restricted to systems 
containing just pions and nucleons and considered basically
the first five processes given in Fig.1 (A)
\cite{OK92,CPS92,FC94,RR94,B94,RR95}. These processes 
constitute an autonomous chiral family, incorporate 
correctly the well known cancellations of the intermediate
$\pi$N amplitude \cite{BR96,BRR97} but correspond to an
intermediate amplitude which is too simple for
reproducing $\pi$N experimental data
\cite{H83}. 

The extension of this minimal model so as to
include other degrees of freedom was considered by 
Ord\'o\~nez, Ray and van Kolck \cite{ORK94,ORK96} and
by ourselves \cite{R95,RR97}. In the former case, a very
general effective Lagrangian was used, which included 
explicitly the interactions of pions, nucleons and deltas 
and contained some free parameters representing other 
interactions. In principle, these parameters could be  obtained
from other physical processes, but these authors chose to 
adjust them to NN scattering data. Using a non-relativistic 
cut-off of the order of the rho mass, they could achieve 
a qualitative description of all NN observables.

In our approach, the intermediate $\pi$N 
amplitude involved the interactions of pions and nucleons,
supplemented by empirical information in the form of th
H\"ohler, Jacob and Strauss (HJS) coefficients \cite{HJS72}.
In general, the physical amplitude for the process 
$\pi^\alpha(k) N(p) \rightarrow \pi^\beta(k') N(p')$
may be described by two independent variables,
$\nu = \frac{1}{4m} (p+p')\cdot(k+k')$ and
$t = (k-k')^2$. In order to obtain the HJS coefficients,
one subtracts the nucleon pole from the empirical $\pi$N
amplitude and uses dispersion relations to continue analytically
the remainder (R) to an unphysical region around the point
$\nu=0$, $t=0$. The HJS coefficients are then obtained by
expanding this remainder in a power series in $\nu$
and $t$. The use of the subthreshold coefficients is particularly
suited to the calculation of the TPEP at large distances, since 
this part of the potential is determined by
the intermediate $\pi$N amplitude in the very neighbourhood of the
point $\nu = 0$, $t =0$. Thus, in this aspect, our approach is very different
from that of Ord\'o\~nez, Ray, and van Kolck.

In our calculation, the TPEP was derived from the ten diagrams
depicted in Fig.1, representing both interactions involving only
nucleons (A) and other degrees of freedom (B).
The main features of the asymptotic TPEP were extensively 
discussed in \cite{RR97} and here we just recall some of the 
conclusions of that paper. One of them is that the scalar-isoscalar
component of the TPEP at large distances is attractive and therefore 
qualitatively coherent with a well known feature of the nuclear
force. As far as dynamics is concerned, we have shown that chiral 
symmetry is responsible for large cancellations in the pure nucleon
sector (Fig.1, A) \cite{BR96,BRR97} and eventually the 
contributions from this sector turn out to be much smaller than 
those arising from other degrees of freedom. 
The main contribution to the intermediate attraction is due to 
processes containing nucleons in one leg and the remainder degrees of freedom
in the other. 

The fact that our calculation of the TPEP did not
contain free parameters means that it yields predictions for NN
observables, whose study is the main goal of the present work.
We assume that, for distances larger than 2.5 fm, the NN interaction
is given by just the OPEP and the TPEP calculated in ref. \cite{RR97}
and try to determine the values of the angular momentum and the energy 
regions for which observables can be ascribed to these components only.
Our presentation is divided as follows: in Sec.II we discuss our 
method of work and in Sec.III we give our results and conclusions.
        

\section{DYNAMICS}

In general, it is not easy to isolate unambiguously the observables
associated with a particular region of a given potential.
Nevertheless, in many cases, the centrifugal barrier can suppress
a significant part of the short range interaction  
and one is left only with contributions from the tail 
of the force. For instance, in a study of the influence of the 
OPEP over NN observables, we have obtained, for most 
waves with $\ell > 2$, purely pionic
phase shifts and mixing parameters, which did not depend on the
short range features of the interaction \cite{BR94}. In the case of
the OPEP, this kind of result is possible because the potential is 
not too strong.

In the present problem, even if one is willing to consider the
influence of the TPEP only for separations larger than 2.5 fm, 
where the results of Ref. \cite{RR97} are mathematically
reliable, one
needs to use expressions which are valid for all
distances.
As it is important have close control of the regions of the 
potential that contribute to the observables, 
we employ the so called variable phase 
method. It is fully equivalent to the Schr\"odinger equation 
and provides a clear spatial picture of the way phase shifts and 
mixing parameters are structured. For the sake of completeness, we
summarize here the main equations used in our calculation. 
In the case of uncoupled channels, the wave function $u_J(r)$ 
with angular momentum $J$ is written as \cite{C63}

\begin{equation}
u_J(r) = c_J(k,r) \; \hat{j}_J(kr) - s_J(k,r) \; \hat{n}_J(kr) \;,
\end{equation}
\label{1}

\noindent
where $\hat{j}_J$ and $\hat{n}_J$ are the usual Bessel and Neumann 
functions multiplied by $kr$. The functions $c_J$ and $s_J$,
for a potential $V_J(r)$, are given by

\begin{equation}
c_J(k,r) = 1 -\frac{m}{k} \int^r_0 \;d\rho 
\;V_J(\rho) \;\hat{j}_J(k\rho)\;u_J(\rho) \;,
\end{equation}
\label{2}

\begin{equation}
s_J(k,r) = -\frac{m}{k} \int^r_0 \;d\rho 
\;V_J(\rho) \;\hat{n}_J(k\rho)\;u_J(\rho) \;.
\end{equation}
\label{3}

\noindent
The variable phase $D_J(k,r)$ is defined as

\begin{equation}
D_J = tan^{-1} (\frac{s_J}{c_J}) \;
\end{equation}
\label{4}

\noindent
and, by construction, it vanishes at the origin and yields the observable
phase shift $\delta_J$ when $r$ tends to infinity. Differentiating 
(2) and (3), using (1), and manipulating the result, one obtains the 
differential equation

\begin{equation}
D'_J = - \frac{m}{k}\; V_J\; P^2_J(D_j) \;,
\end{equation}
\label{5}

\noindent
where $D'_J = \frac{dD_J}{dr}$ and the structure function $P_J$ is
given by 

\begin{equation}
P_J = \hat{j}_J\;cos(D_J) - \hat{n}_J\;sin(D_J) \;.
\end{equation}
\label{6}

In the case of coupled channels, one has two phases, $D_{Jm}$ and
$D_{Jp}$, with $m=L-1$ and $p=L+1$, and a mixing parameter $E_J$, 
that depend on $r$ and become the observables $\delta_{Jm}$, 
$\delta_{Jp}$ and $\epsilon_J$ when r tends to infinity. Denoting
the diagonal and tensor components of the potential by $W_{JL}$ and
$T_J$, one has the following coupled differential equations \cite{B67}

\begin{eqnarray}
D'_{Jm} = &-& \frac{m}{k\;cos(2E_J)}
\left\{ W_{Jm} \left[ cos^4(E_J)\;P_m^2 - sin^4(E_J)\;Q_m^2\right]
\right.
\nonumber\\
&-& \left.
W_{Jp}\;sin^2(E_J)\;cos^2(E_J)\;(P_p^2 - Q_p^2)
\right.
\nonumber\\
&-& \left.
2\;T_J\;sin(E_J)\;cos(E_J)\left[ cos^2(E_J)\;P_m\;Q_p
-sin^2(E_J)\;P_p\;Q_m\right] \right\} \;,
\end{eqnarray}
\label{7}

\begin{eqnarray}
E'_{Jm} = &-& \frac{m}{k}
\left\{ T_J \left[ cos^2(E_J)\;P_m\;P_p + sin^2(E_J)\;Q_m\;Q_p\right]
\right.
\nonumber\\
&-& \left.
W_{Jm}\;sin(E_J)\;cos(E_J)\;P_m\;Q_m
- W_{Jp}\;sin(E_J)\;cos(E_J)\;P_p\;Q_p \right\} \;.
\end{eqnarray}
\label{8}

In these expressions the structure functions $P_L$ and $Q_L$
are defined as

\begin{eqnarray}
P_L &=& \hat{j}_L\;cos(D_{JL}) - \hat{n}_L\;sin(D_{JL}) \;,
\label{9}\\
Q_L &=& \hat{j}_L\;sin(D_{JL}) + \hat{n}_L\;cos(D_{JL}) \;.
\end{eqnarray}
\label{10}

\noindent
The equation for $D_{Jp}$ is obtained by exchanging the labels
$m$ and $p$ in (7).

As far as the interaction is concerned, we consider just the OPEP 
$(V_\pi)$ and
the TPEP, which are assumed to represent the full potential for
distances greater than 2.5 fm.
As pointed out in the introduction, the TPEP is determined
by two kinds of contributions, one generated in the pure pion-nucleon
sector $(V_N)$ and the other associated with the remainder degrees 
of freedom $(V_R)$.
The direct inspection of these potentials indicates that the former 
is comparable to the OPEP, whereas the latter is rather strong
for $r > 2.0$ fm.
In order to calculate the observables, one needs to regularize the various 
potentials at short distances. In the case of OPEP, the 
regularization is achieved by cutting it at a radius
$r_\pi$ and replacing the inner part by the constant value 
$V_\pi(r_\pi)$. As $V_N$ is comparable to the OPEP, we adopt the same 
regularization procedure for it, with a radius $r_N$. 

The regularization of $V_R$ is more problematic. For distances 
around 1.0 fm, the value of the cental component of the potential 
is about $-25$ GeV. On the other hand, in Ref.\cite{RR97} we
have argued that the asymptotic TPEP is mathematically reliable only
for distances larger than 2.5 fm, indicating that the odd behaviour of
the TPEP at short distances is unphysical and associated with the use 
of equations outside their domain of validity.
Inspecting the equations used in that
work, it is easy to relate this behaviour to the HJS coefficients
involving high powers of $\nu$ and $t$ in the intermediate $\pi$N
amplitude. On the other hand, restricting ourselves to the first two
leading contributions, due to the coefficients of the terms
$\nu^0 t^0$ and $\nu^0 t^1$, we keep most of the asymptotic
potential and get values around $-150$ MeV in the neighbourhood of
1.0 fm, which are still large, but much more reasonable. Therefore
we base the present study on this leading potential, which is 
regularized by means of a step function $\theta(r_R)$.

In this calculation one can rely only on those results which are 
independent of the radii used in the
regularization procedure. In order to control this independence,
we adopt $r_N = r_\pi$, vary $r_\pi$ in the interval
$0.8-1.0$ fm and discard the cases where the contribution
of $V_\pi + V_N$ to the observables 
varies more than $1.0\%$. Concerning $V_R$, the
preceding discussion suggests that one should be interest in effects
due to the region $r>2.5$ fm and hence we consider values of $r_R$ between
1.5 fm and 2.5 fm and study the effect of this variation over the observables.
This produces an indication of both the stability
of the results and the importance of the inner part of the potential.
When the variation of the results is less than $5\%$, we take them 
as predictions of the potential. For larger deviations, we consider them
as estimates.


\section{RESULTS AND CONCLUSIONS}

We have calculated the predictions for NN observables
produced by a chiral 
potential \cite{RR97} involving
only the exchanges of one and two pions, assumed to
represent the full interaction for distances larger than 2.5 fm.
Since we are interested only in the cases where 
the centrifugal barrier cuts naturally the
inner parts of the force, we used the variable phase method 
to control this aspect of the problem. 
One may acquire a feeling for this method by looking at  
Fig. 2, where we display the variable phases for the wave $^1G_4$, divided
by their asymptotic values, for three different energies.
It is possible to see that, as expected, higher energies probe more the 
interior of the potential. It is also interesting to note 
that the variable phase method allows one to make quantitative 
statements such as, for instance, the radius for which the phase
attains a given percentage of its final value.

In the case of uncoupled waves, the interplay between the centrifugal 
barrier and the regularization of the background is rather
intuitive, but this is not the case of coupled systems. In order to
clarify this point, we show in Fig. 3 the variable phases for the
$^3D_3-^3G_3$ system for $E=200$ MeV, due just to the 
background and regularized at either
at 0.7 fm or 1.0 fm. One sees that the $^3D_3$ curves
depend strongly on the cutoff, whereas those describing
$\epsilon_3$ and $^3G_3$ are very stable and
cannot be distinguished with the naked eye. 
The wave $^3G_3$ is negligible for distances smaller than
2.0 fm due to the centrifugal barrier, meaning that the system is
effectively uncoupled up to that distance. The construction of $^3D_3$
extends up to 3.5 fm, where $\epsilon_3$ is maximum, but the two 
other observables become asymptotic much later, around 5.0 fm. 
This indicates that, even for coupled waves, the various components get 
their contributions from different regions in space.

We found out that the observables associated with the 
following waves depend more than $1 \%$ on the cutoff used for 
the $V_\pi+V_N$ background and hence are not 
suited for our study: $^1S_0$, $^3P_0$,
$^3P_2$, $^3S_1, \epsilon_1, ^3D_1$ and $^3D_3$. 

In Figs. 4-7, we display our results for the phase shifts as a function of 
the laboratory energy. They include predictions from 
the OPEP cut at 1.0 fm, the
sum of the OPEP and $V_N$
cut at 1.0 fm  and the total results produced by cutting
the contribution of $V_R$ at 1.5 fm ($\chi(1.5)$) 
and 2.5 fm ($\chi(2.5)$), and the experimental 
values are taken from SAID  VZ40 solution \cite{SAID}. For comparative
purposes, we also include the predictions of the Argonne
\cite{A84} phenomenological potential.

The dominant part of the chiral TPEP is associated with the exchange of a 
scalar-isoscalar system and hence its significance depends strongly on the 
NN channel considered. Therefore in the sequence we comment the main 
features of our results for the various subspaces of spin and isospin.

{\bf{(T=1,S=0)}}; Fig. 4: There are no results for the wave $^1S_0$, since 
it cannot be understood as a TPEP superimposed to a pionic background. 
For the waves $^1D_2$ and $^1G_4$ one has predictions for energies up to 
50 Mev and 300 MeV respectively. As expected, 
the TPEP increases the attraction due to the OPEP and  our
results are very close to the experimental values for the potential cut at 2.5
fm.

{\bf{(T=1,S=1)}}; Figs. 5a,b,c,d: The waves $^3P_0$ and $^3P_2$ depend on the pion 
cutoff and were discarded. For the uncoupled waves $^3P_1$, $^3F_3$, and
$^3H_5$ (Fig. 5a) we obtain predictions which extend up to 300 MeV for the last two of
them. Results for the waves $^3P_1$ and  $^3F_3$ are compatible with experiment,
but this does not happen for the $^3H_5$ wave. 
In the case of the
coupled waves, that with lowest orbital angular momentum tend,
as expected, to be much more influenced by the cutoff used for the
OPEP than the other ones. We obtain predictions in all cases, but 
the mixing parameters are heavily dominated by the OPEP
and yield very little information about the TPEP.
For the waves $^3F_2$ (Fig. 5b) and $^3F_4 $ (Fig. 5c), 
the differences with experimental values are small, whereas for the waves 
$^3H_4$ (Fig. 5c) and $^3H_6$ (Fig. 5d) they are important.

{\bf{(T=0,S=0)}}; Fig. 6: Our calculation yields predictions for all the
waves in this sector, namely $^1P_1$, $^1F_3$, and $^1H_5$, 
generally quite close to the pure OPEP ones, reflecting the fact that 
our central potential in this channel is small. Results are also close to
experiment.

{\bf{(T=0,S=1)}}; Figs. 7a,b,c:  The observables $^3S_1,\,\epsilon_1,\,^3D_1$ and $^3D_3$
depend on 
the pion cutoff and are not considered. In all cases, coupled and
uncoupled, our results are dominated by the OPEP and close  to experiment.

In order to assess the general trends of the various observables presented in
Figs. 4-7, it is useful to recall that the relative strengthes of the central
and tensor OPEP in the channels $(T,S)=(1,0),(1,1),(0,0),\text{ and }(0,1)$ are
respectively $1:\frac{1}{3}:3:1$ and $0:\frac{1}{3}:0:1$. This means that one
pion exchange is more important in the channels with $T=0$ and hence the good
agreement between predictions and experimental results noted in Figs. 6 and 7
may be ascribed to OPEP physics.

As far as the channels with $T=1$ are concerned (Figs. 4 and 5), the tensor 
interaction makes the OPEP to be more important for triplet waves, in spite of
its weaker central component. Therefore the role of the TPEP is more evident in
the waves $^1D_2$ and $^1G_4$ (Fig. 4), where chiral predictions agree well with
experiment. In the case of triplet waves (Figs. 5a,b,c,d), one also finds that the chiral
potential is able to reproduce experimental data  when $\ell < 5$, but this
does not happen for $H$ waves. The behavior of these waves are peculiar, since
they have a high orbital angular momentum and hence should be close of being
OPEP dominated. Indeed, our results show that predictions from the chiral
potential for H waves are not far from the OPEP and also depend little on the cutting
radius.  Part of the discrepancies observed may be associated with the fact
that we have used $g^2/4\pi=14.28$ for the $\pi N$ coupling constant
\cite{HJS72} whereas the SAID analysis is based on the value $g^2/4\pi=13.7$.
It is also worth pointing out that there is a 10\% difference between the
experimental $pn$ and $pp$ solutions and the discrepancies would be reduced if
the latter were used. However, even if these factors were considered, the
experimental data would still seem to suggest that the TPEP is repulsive for
the waves $^3H_4$ and $^3H_6$, something which is rather difficult to explain
theoretically.

A general conclusion that can be drawn from the present study concerns the
details of the TPEP. 
As discussed in the introduction and represented in Fig. 1, 
it consists in a sum of terms, arising from both the
pure pion nucleon sector and from interactions involving other 
degrees of freedom. Our results show that
the former contributions are very small, indicating that the 
numerical significance of the TPEP is essentially due to the
interplay between nucleon and other degrees of freedom.

In this work we tested the chiral TPEP derived in Ref.\cite{RR97}, which 
is based on subthreshold $\pi N$ data and contains no free parameters.
Our results have shown that it is rather consistent 
with experiment\footnote{Note added in revision: In a recent work, the
predictions from a similar chiral potential were presented
\protect\cite{KBW97}, which agree qualitatively with those produced here.}.

\section{Acknowledgments}
 
M.R.R would like to thank the kind hospitality of
the Division de Physique Theorique de l'Istitut de Physique Nucleaire,
Orsay, France,
where this work was performed and FAPESP (Brazilian
Agency), for financial support. The work of C.A. da Rocha,
was supported by Grant No. 200154/95-8 from CNPq Brazilian agency.
This work was partially supported by U.S. Department of Energy.



\begin{figure}
\epsfxsize=13.0cm
\epsfysize=5.5cm
\centerline{\epsffile{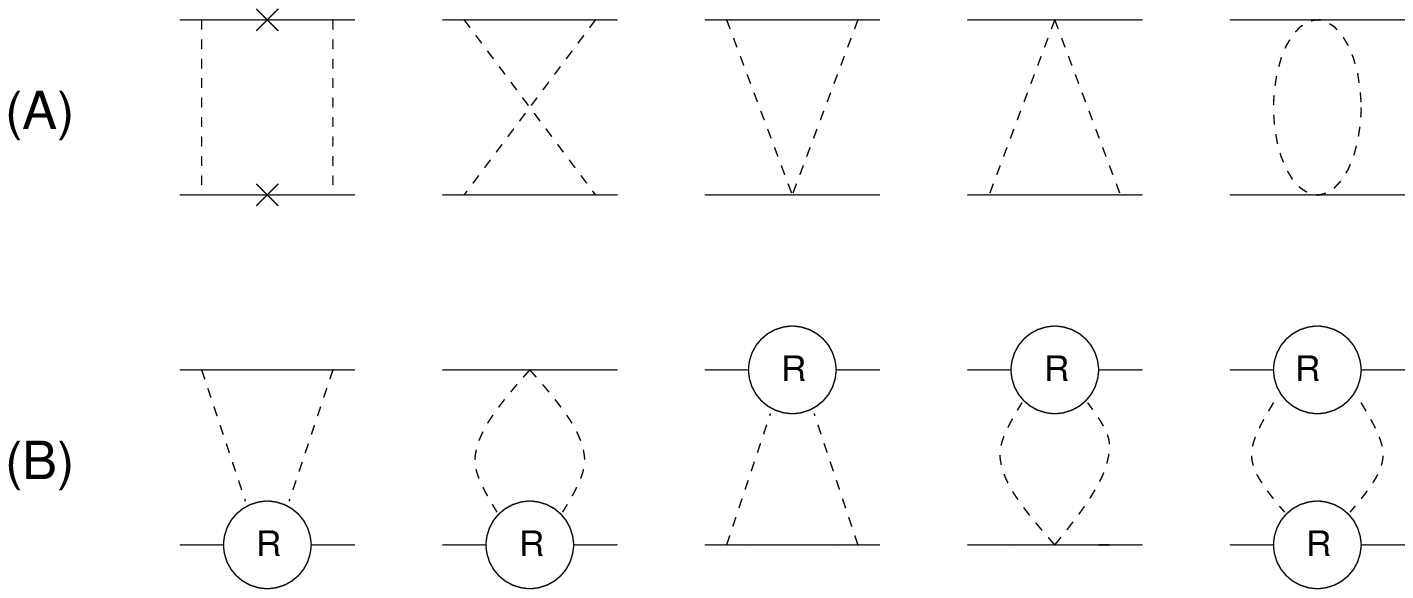}}
\vspace{1.0cm}
\caption{Contributions to the two-pion exchange nucleon-nucleon potential
from the pure pion-nucleon sector (A) and for processes involving
other degrees of freedom (B). The crosses in the nucleon lines of the 
first diagram indicate that the iterated OPEP was subtracted.}
\label{Fig.1}
\end{figure}

\newpage
\newpage
\begin{figure}
\epsfxsize=13.0cm
\epsfysize=13.0cm
\centerline{\epsffile{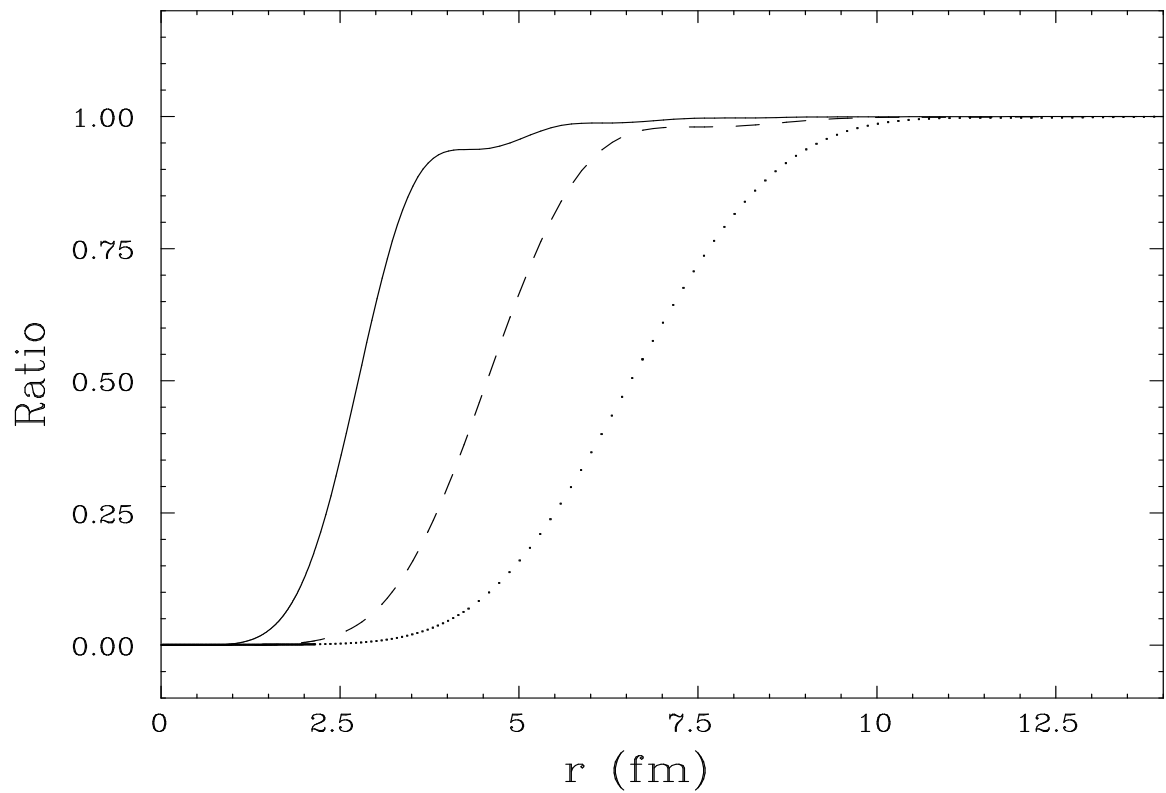}}
\vspace{1.0cm}
\caption{Variable phases (normalized to 1) for the wave $^1G_4$ at the 
energies $E=40$ MeV (dotted line), $E=100$ MeV (dashed line) and
$E=300$ MeV (continuous line).}
\label{Fig.2}    
\end{figure}

\newpage
\begin{figure}
\epsfxsize=13.0cm
\epsfysize=13.0cm
\centerline{\epsffile{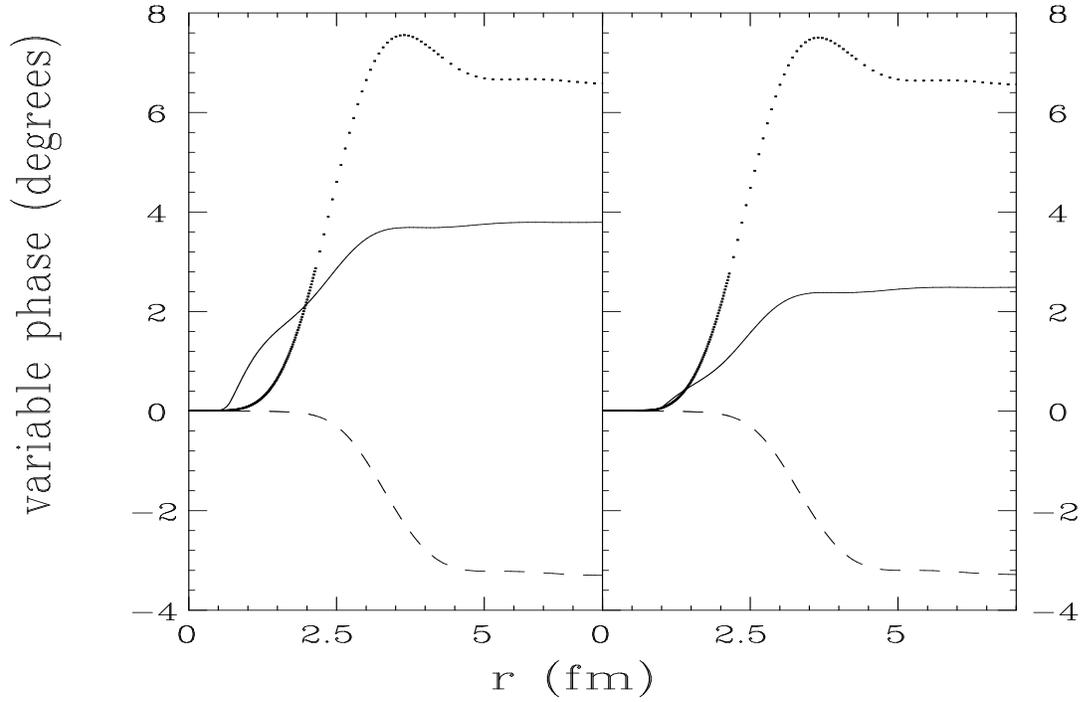}}
\vspace{1.0cm}
\caption{Variable phases for the waves $^3D_3$ (dotted line),
$^3G_3$ (dashed line) and the mixing parameter $\epsilon_3$
(solid line) for background potentials cut at 0.7 fm (left) 
and 1.0 fm (right).}
\label{Fig.3}
\end{figure}

\newpage
\begin{figure}
\epsfxsize=15.0cm
\epsfysize=20.0cm
\centerline{\epsffile{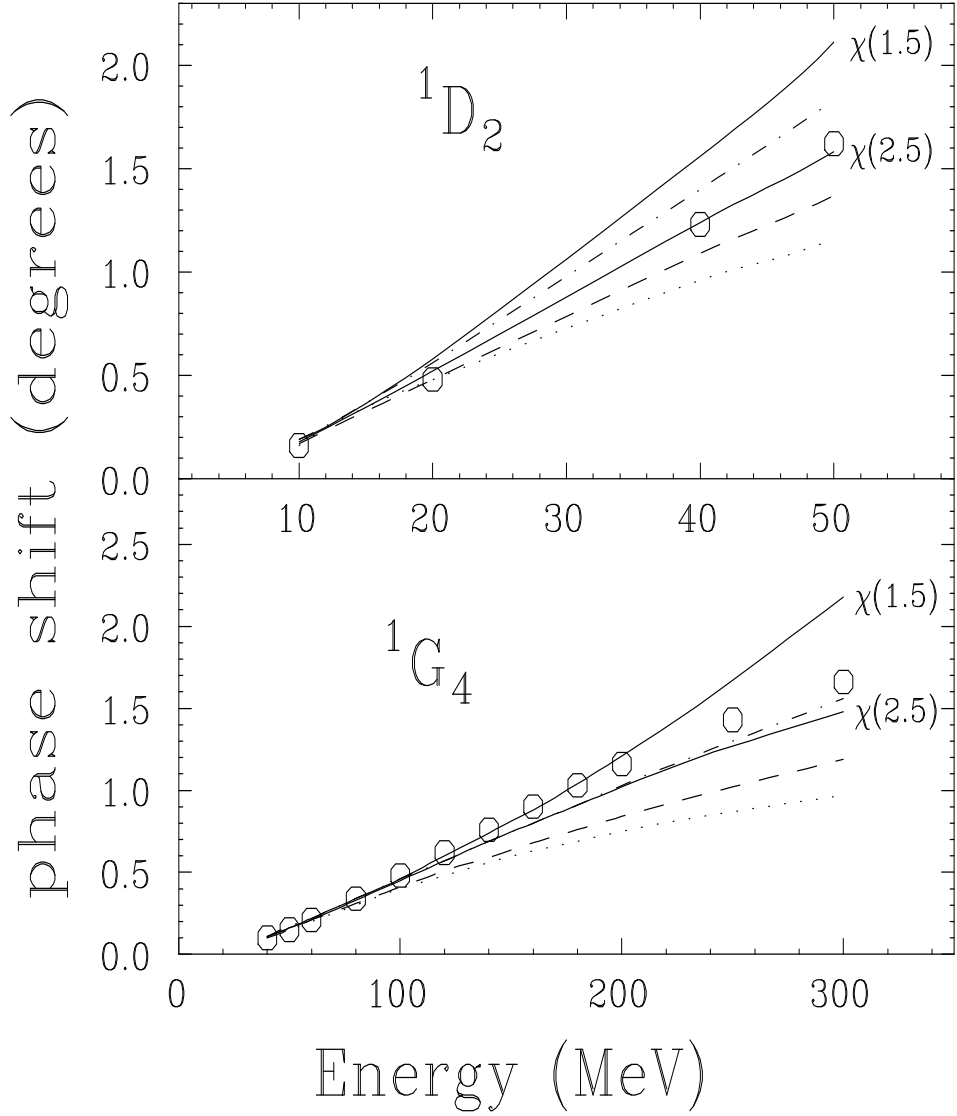}}
\vspace{-2cm}
\caption{Phase shifts for the channel $(T=1,S=0)$ due to the OPEP cut at 1.0 fm
(dotted line), the sum of the OPEP and the pure pion-nucleon TPEP cut at 1.0 fm
(dashed line) and the full chiral potential cut at 1.5 fm and 2.5 fm
(continuous lines labeled $\chi (1.5)$ and $\chi (2.5)$). For comparative
purposes, the predictions of the argonne v14 potential are also included
(dashed-dotted line). The experimental points represent the $pn$ solution of
SAID \protect\cite{SAID}.}
\label{Fig.4}
\end{figure}

\newpage
\begin{figure}
\epsfxsize=8.0cm
\epsfysize=12cm
\epsffile{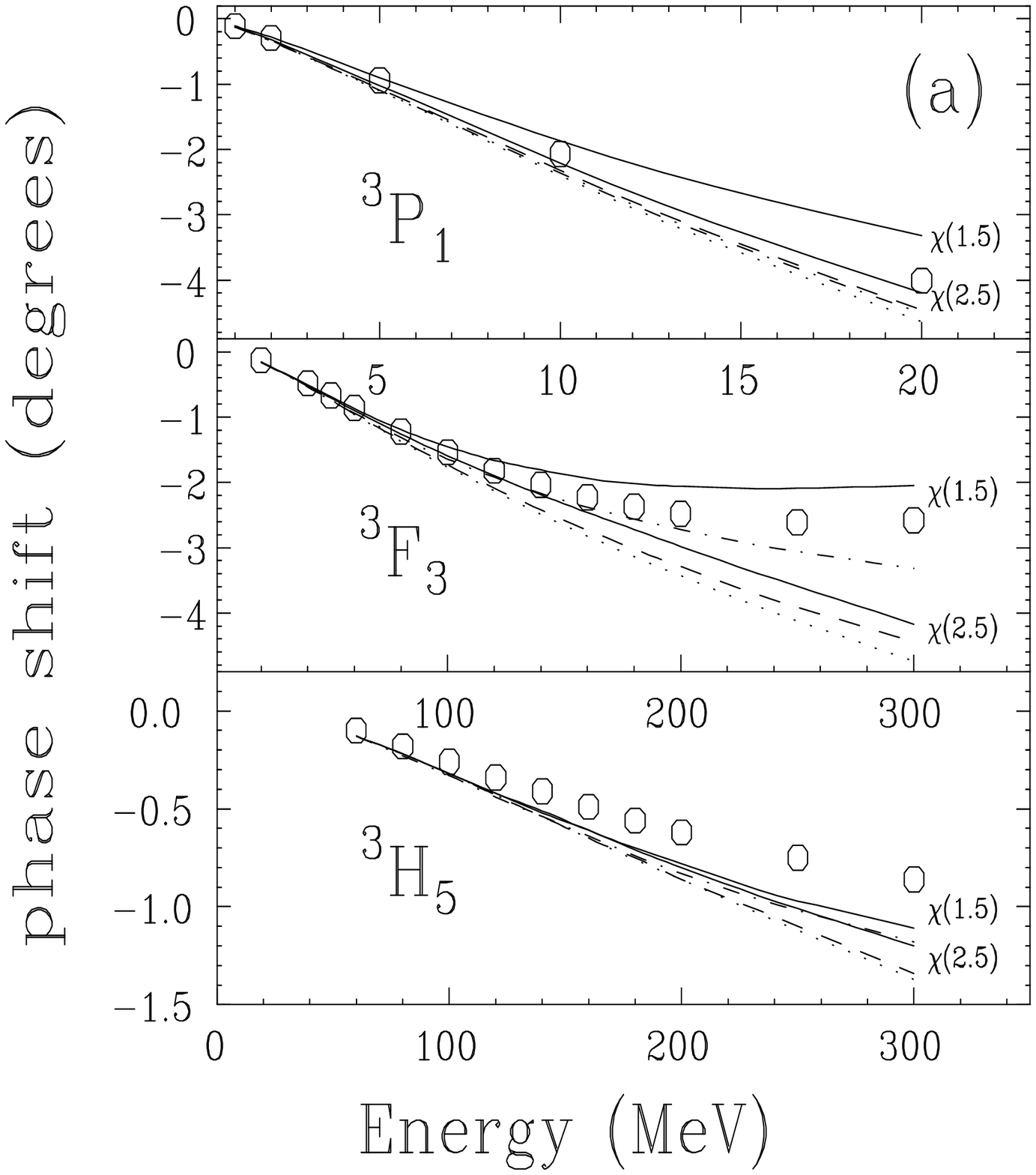}
\vspace{-2.0cm}
\epsfxsize=8.0cm
\epsfysize=12cm
\epsffile{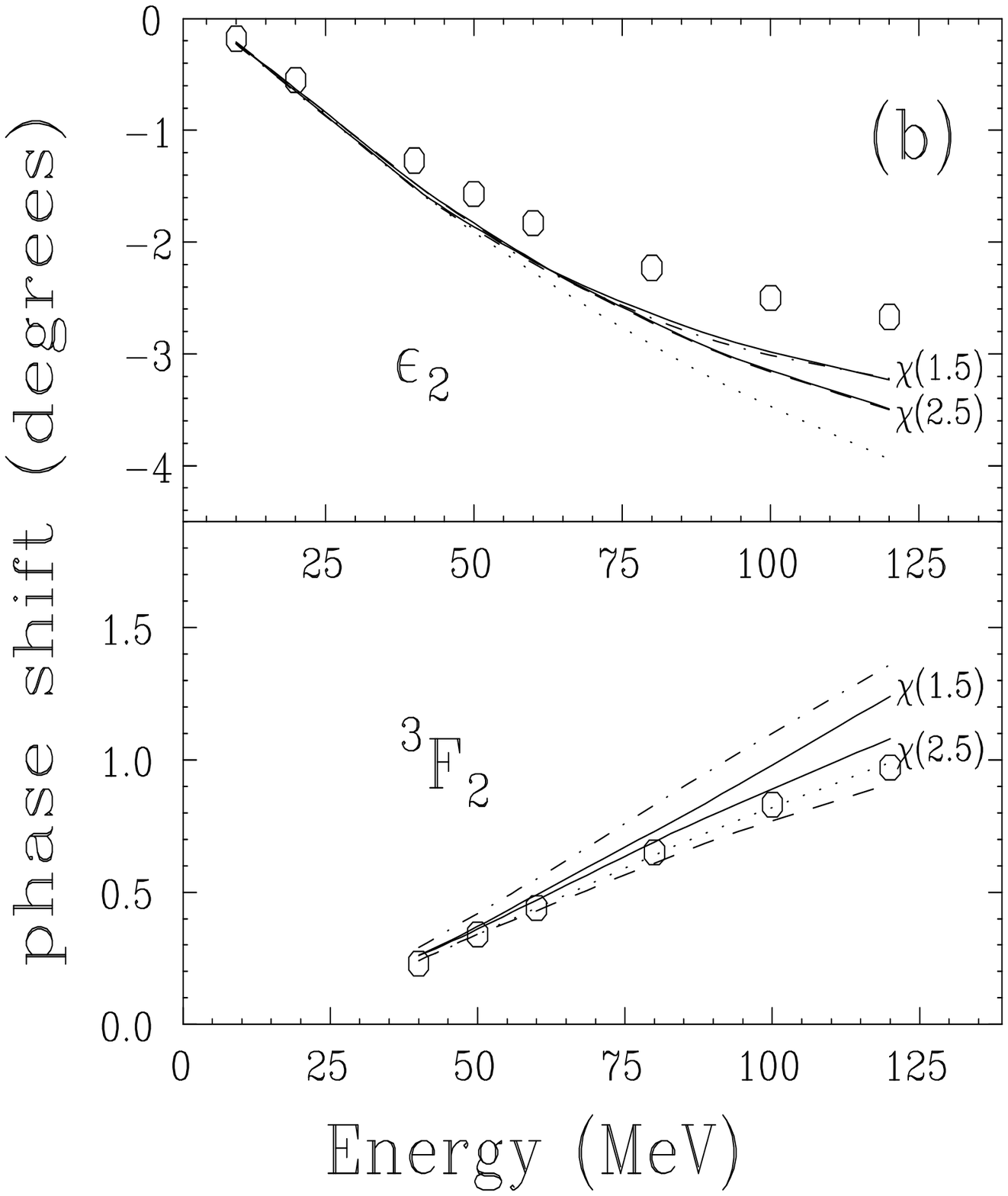}
\vspace{-22cm}
\epsfxsize=8.0cm
\epsfysize=12cm
\rightline{\epsffile{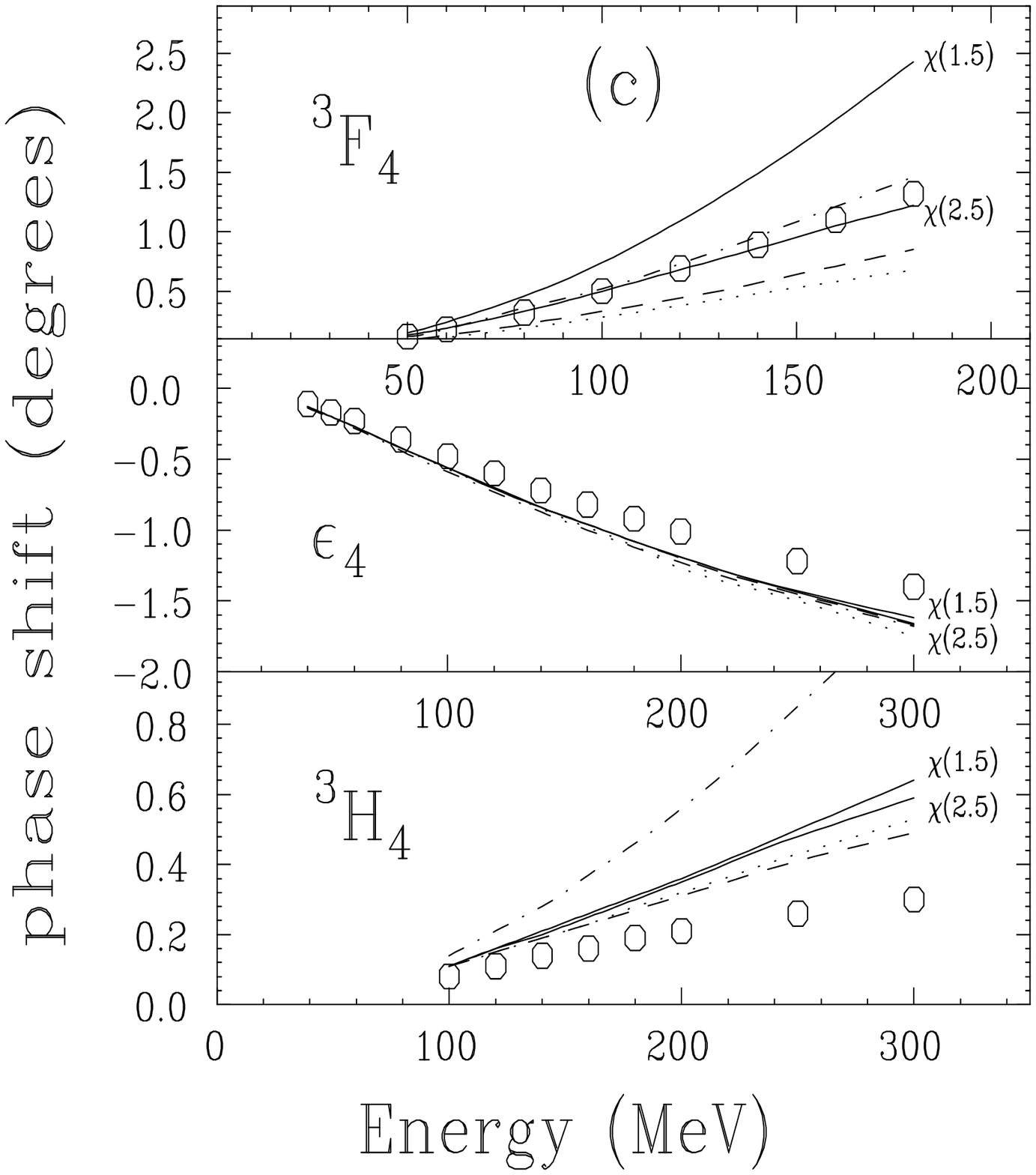}}
\vspace{-2.0cm}
\epsfxsize=8.0cm
\epsfysize=12cm
\rightline{\epsffile{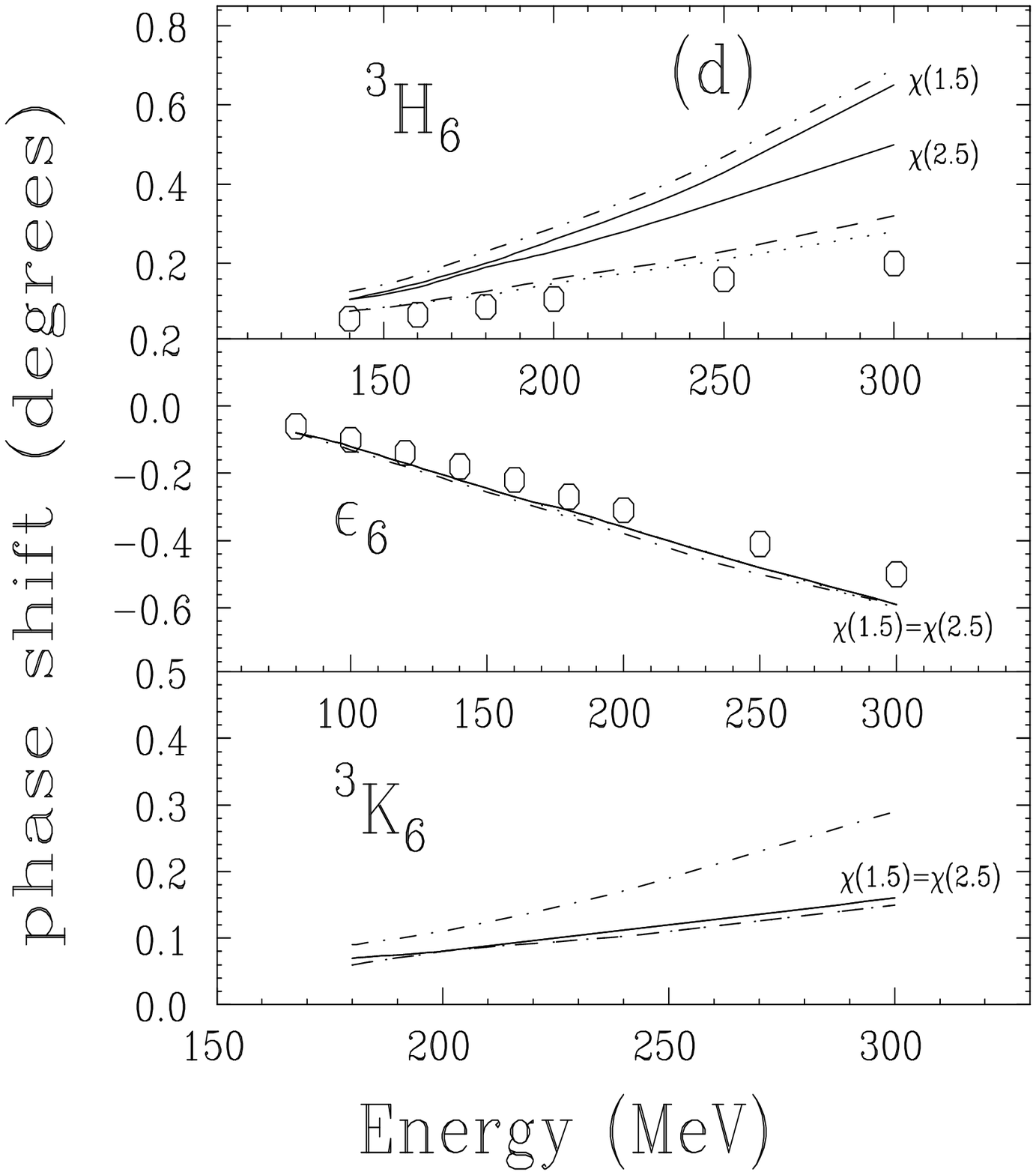}}
\vspace{-1.5cm}
\caption{Phase shifts for the channel $(T=1,S=1)$; (a) uncoupled waves; (b)
$J=2$ coupled channel; (c) $J=4$ coupled channel; (d) $J=6$ coupled channel.
Conventions are given in Fig. 4.}
\label{Fig.5}
\end{figure}

\newpage
\begin{figure}
\epsfxsize=15.0cm
\epsfysize=22.0cm
\centerline{\epsffile{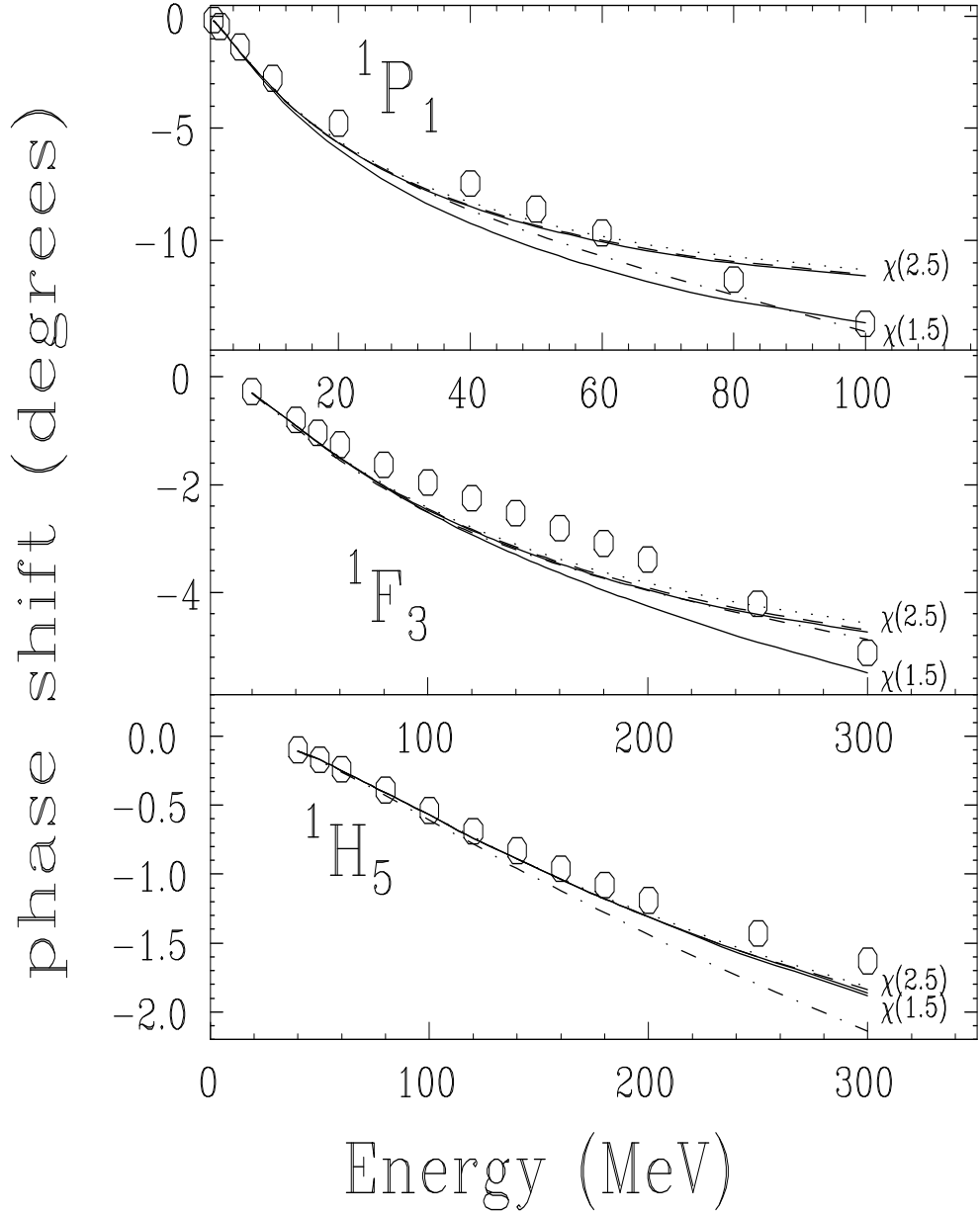}}
\vspace{-2cm}
\caption{Phase shifts for the channel $(T=0,S=0)$; conventions are given in Fig. 4.}
\label{Fig.6}
\end{figure}

\newpage
\begin{figure}
\epsfxsize=8.0cm
\epsfysize=12cm
\epsffile{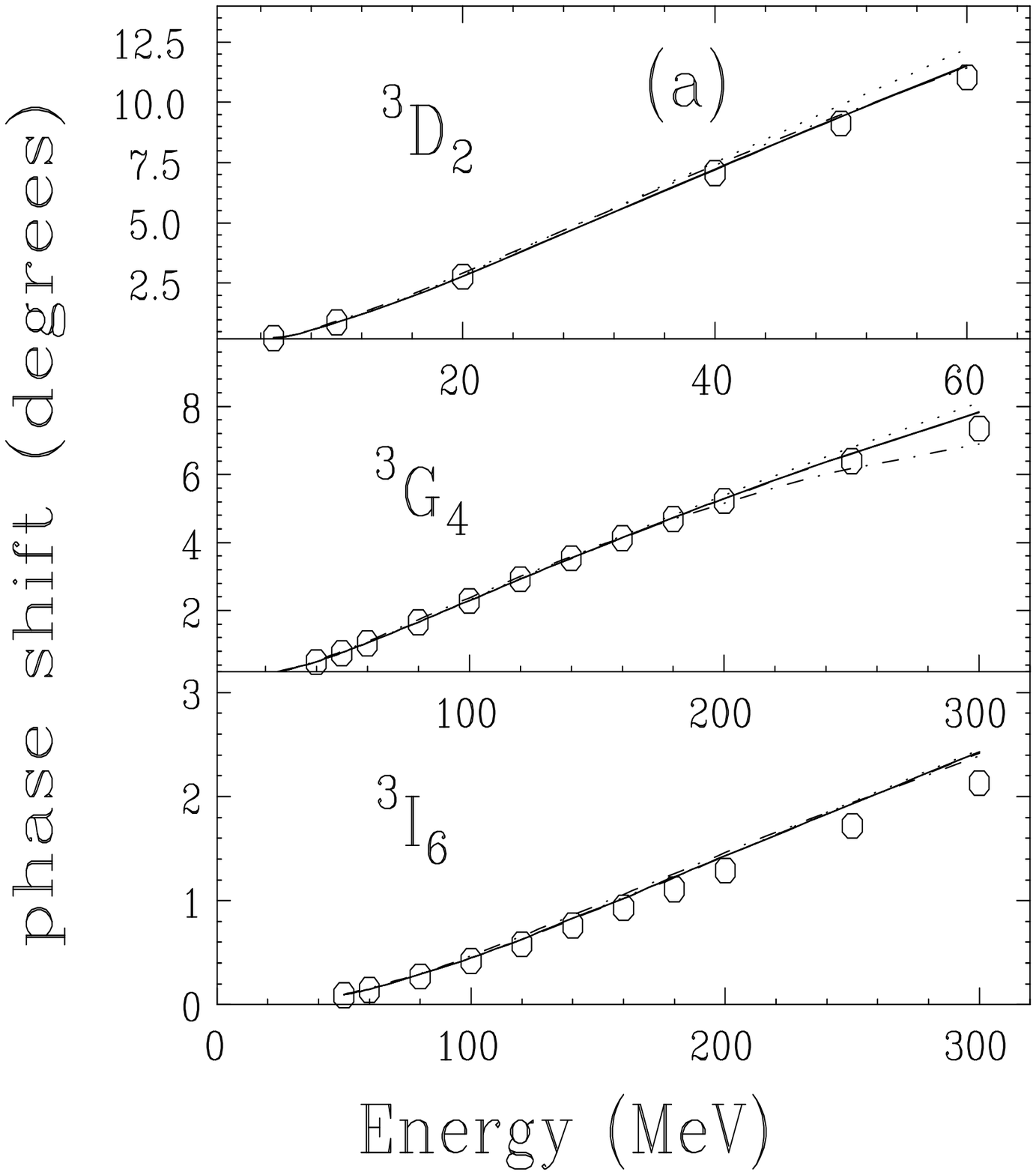}
\vspace{-2.0cm}
\epsfxsize=8.0cm
\epsfysize=12cm
\epsffile{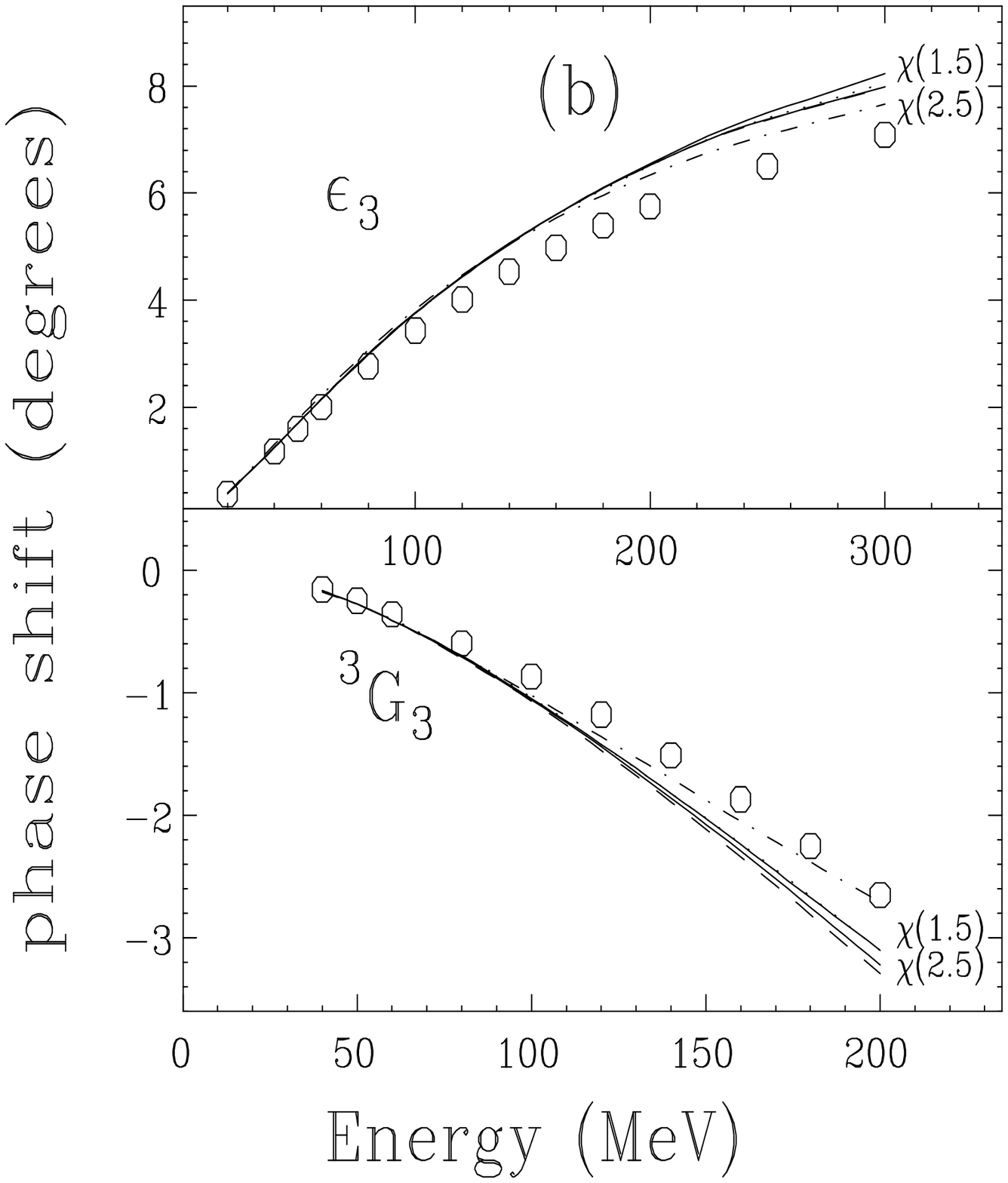}
\vspace{-22cm}
\epsfxsize=8.0cm
\epsfysize=12cm
\rightline{\epsffile{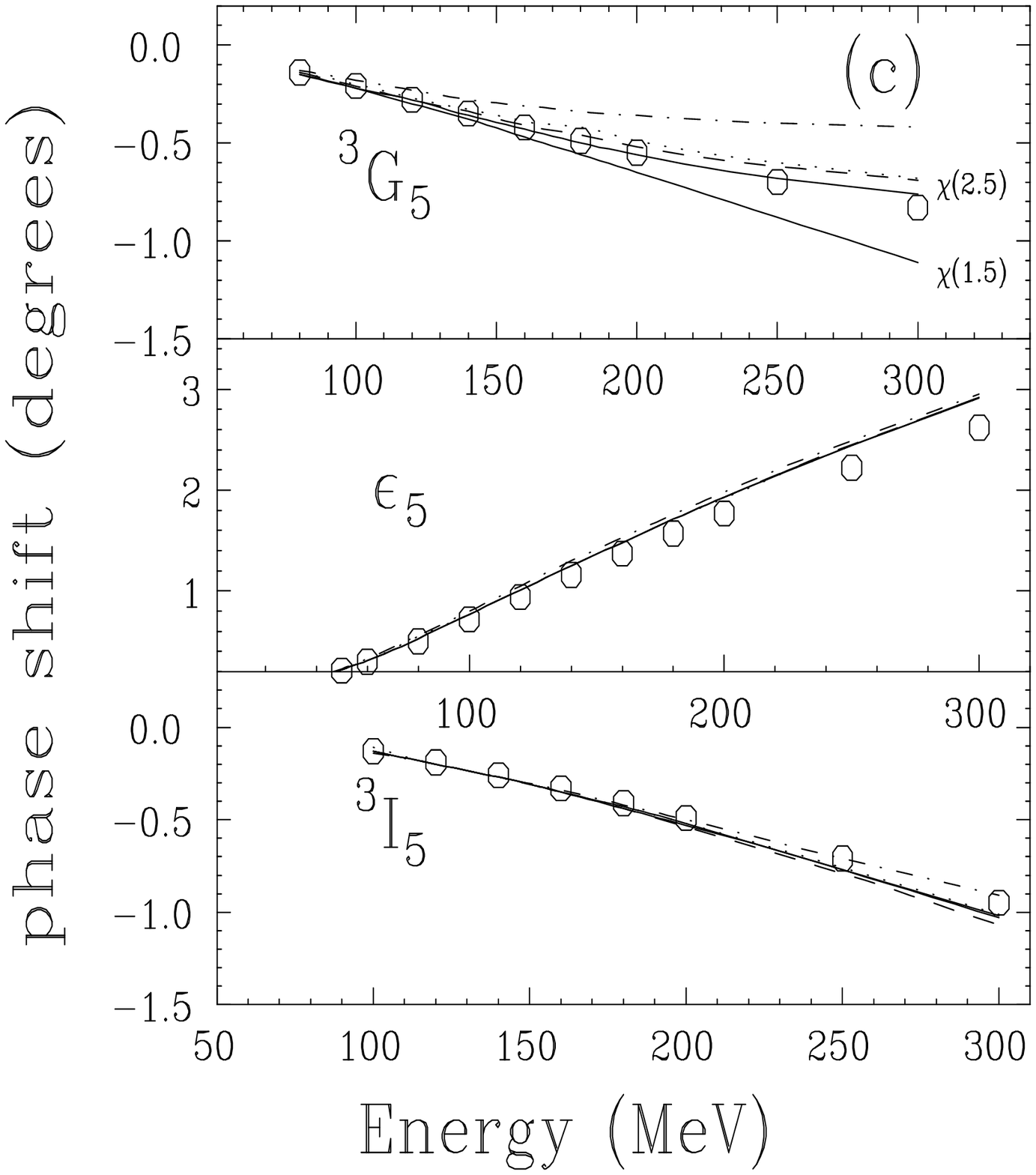}}
\vspace{9cm}
\caption{Phase shifts for the channel $(T=1,S=1)$; (a) uncoupled waves; (b)
$J=3$ coupled channel; (c) $J=5$ coupled channel.
Conventions are given in Fig. 4.}
\label{Fig.7}
\end{figure}

\end{document}